\documentclass[a4paper]{PoS}

\title{Direct photon production in relativistic heavy-ion collisions -- a theory update}

\ShortTitle{Direct photon production in relativistic heavy-ion collisions}

\author{\speaker{Charles Gale}\\
        Department of Physics, McGill University, 3600 rue University, Montreal, QC, Canada H3A 2T8\\
        E-mail: \email{gale@physics.mcgill.ca}}

\abstract{For tomographic studies of relativistic nuclear collisions and of the quark-gluon plasma, photons (real and virtual) are unique. They are the only probes than can be both {\it soft} and {\it penetrating}. First we report  on advances in modelling the hadron dynamics of heavy-ion collisions using a hybrid approach which consists of IP-Glasma, relativistic fluid dynamics, and hadronic cascade components. We briefly discuss the ``photon flow puzzle'', and then focus on a recent development in the theory of photon emission from a non-equilibrium, strongly interacting medium.  } 

\FullConference{
12th International Workshop on High-pT Physics in the RHIC/LHC Era\\
		2-5 October, 2017\\
		University of Bergen, Bergen, Norway}

\begin{document}

\section{Introduction}
The study of heavy-ion collisions has revealed the existence of an exotic state of matter: the quark-gluon plasma (QGP). The challenge is now to {\it characterize} the strongly interacting medium through tomographic studies (using jets and other high-$p_T$probes) and global analyses. Electromagnetic observables constitute  ideal and necessary complements to measurements of hadrons; their feeble interaction enables them to freely escape the system, once formed. They therefore have the ability to report on the entire space-time evolution of the colliding system.  Because of this, photons enjoy a unique status: they can be {\em soft}, yet {\em penetrating}. There are no other observables that can make this dual claim.  

In this proceedings contribution, we will report mainly on some of the progress realized in the calculation of soft photons ({\it i.e.} photons with transverse momentum less than a few GeV/c). In this energy window, several sources of photons exist \cite{Gale:2009gc}. We  mention here only two: real prompt photons and real thermal photons, and only really discuss the latter. We start with a necessary assessment of the dynamics of soft to intermediate-energy hadrons.

\section{Hadrons}
The presence of a quark-gluon plasma  (QGP) has been made manifest mostly through the observation of the large final-state collectivity of the measured hadrons, together with the interpretation of those measurements using relativistic fluid dynamics \cite{Gale:2013da}. More specifically, the final momentum distribution of an average number of particle, $N$, in a given event is typically characterized by the coefficients of its Fourier expansion in azimutal angle $\phi$, with complex coefficients, $V_n$. These are then written down as a magnitude $v_n$ and a phase, $n \psi_n$, such that $V_n = \langle e^{i n \phi} \rangle = v_n\, e^{i n \psi_n}$.  The average $\langle \ \rangle$ is taken over particles of a given kind in an event.
\begin{eqnarray}
\frac{d N}{d \phi} = \frac{N}{2 \pi} \sum_{n= - \infty}^{\infty} V_n e^{- i n \phi} = \frac{N}{2 \pi} \left[1 + \sum_{n = 1}^{\infty} 2 v_n \cos n (\phi - \psi_n )\right]
\end{eqnarray}
The second equality is an alternative form,  where $\psi_n$ is a reference angle defined such that $\langle \sin n (\phi - \psi_n) \rangle = 0$.  The flow coefficients $v_n$ can be generalized to include a dependence on $p_T$ and $\eta$, where $p_T$ is the momentum transverse to the beam axis, and $\eta$ is the pseudorapidity \cite{Poskanzer:1998yz,Luzum:2013yya,Gale:2013da}. 
\begin{figure*}[h!]
\centering
%\begin{tabular}{ccc}
\includegraphics[width=0.350\linewidth]{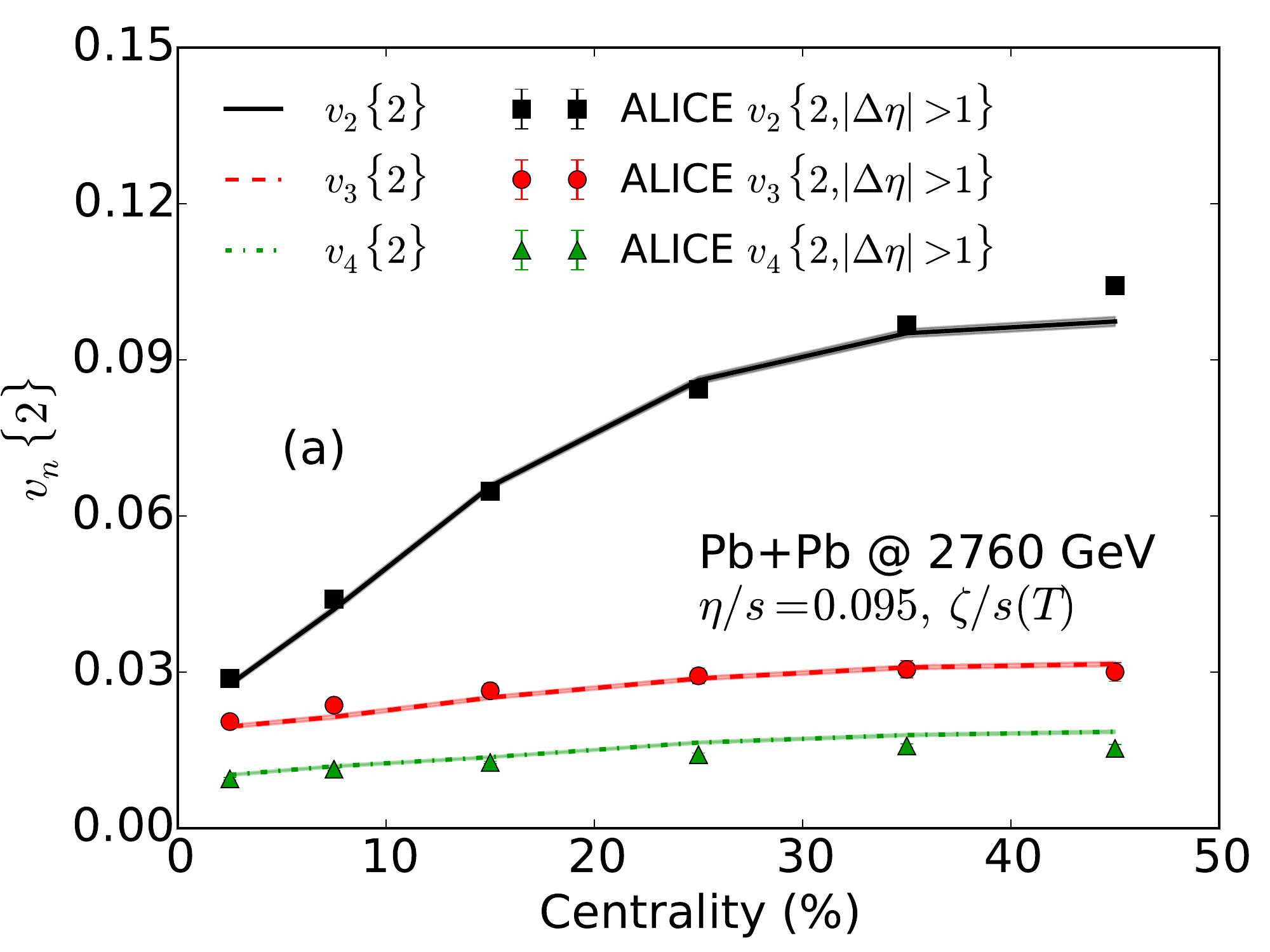} 
\includegraphics[width=0.350\linewidth]{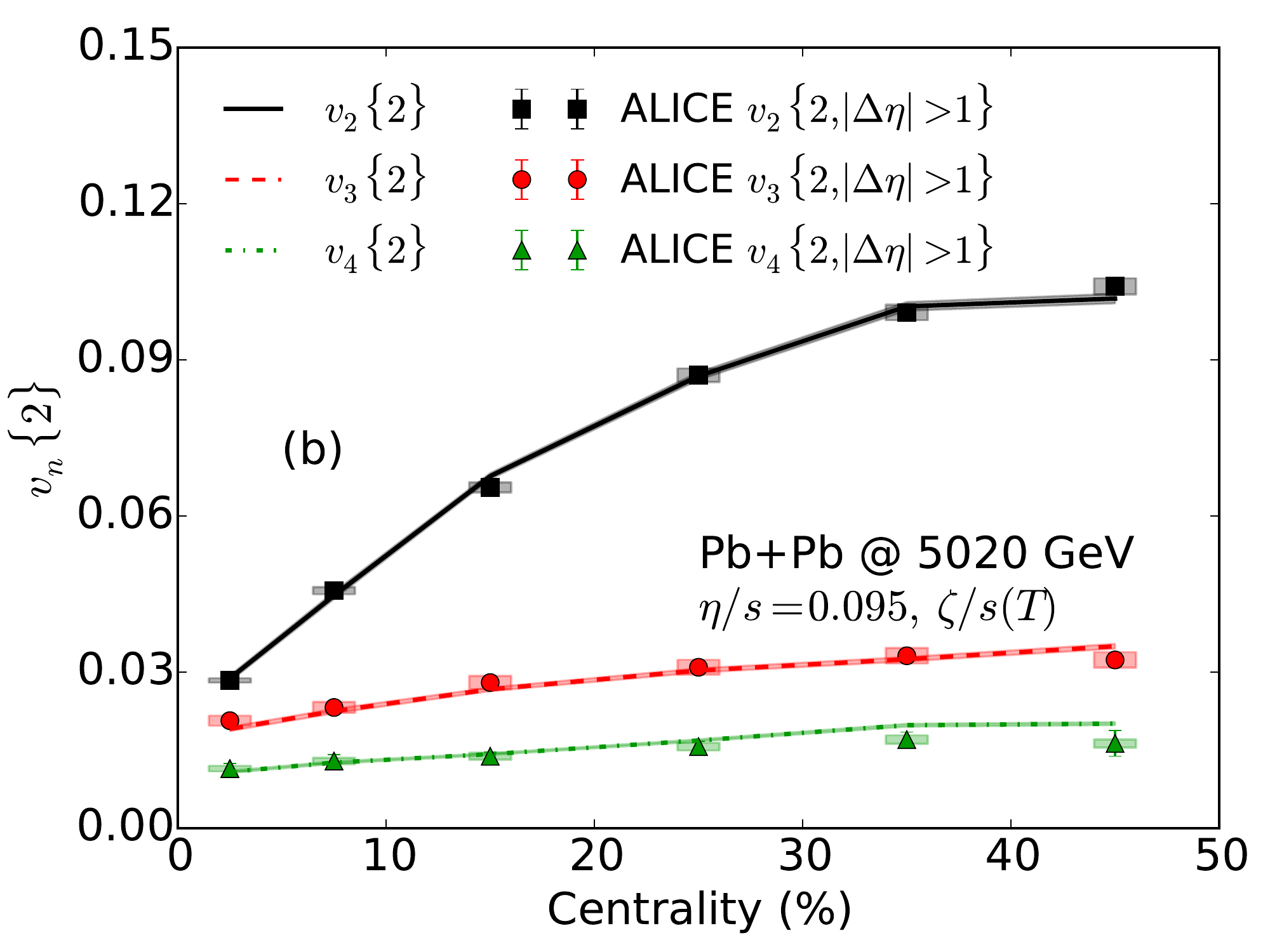} 
%  \includegraphics[width=0.30\linewidth]{Fig1b.pdf} & 
%  \includegraphics[width=0.30\linewidth]{Fig1c.pdf}
%\end{tabular}
\caption{The centrality dependence of the charged hadron integrated flow coefficients for Pb + Pb collisions at 2.76 TeV ({\it Panel (a)}) and at 5.02 TeV ({\it Panel (b)}). The theory (solid curves) is compared against results from the ALICE Collaboration. The plot is from Ref. \cite{McDonald:2016vlt}, and the data are from Ref. \cite{ALICE_flow}.}
\label{int_flow}
\end{figure*}

On the theory side, the interpretation of such measurements have put into light the remarkable effectiveness of relativistic hydrodynamics as a modelling approach to the time-evolution of the strongly interacting system formed during the relativistic collisions of nuclei \cite{Gale:2013da,Jeon:2015dfa}. As an example of recent progress, an integrated state-of-the-art approach \cite{McDonald:2016vlt}  uses as a  representation of  the initial state the IP-Glasma model \cite{IPGlasma}, followed by a hydrodynamical evolution which considers both shear and bulk viscosity coefficients, and then allows a dynamical freeze-out by using hadronic transport in the final stages. The fluid-dynamical approach used in the hybrid model results shown here is {\textsc music}\footnote{http://www.physics.mcgill.ca/music/}. The hybrid approach was recently used to interpret LHC results for Pb + Pb collisions at $\sqrt{s} = 2.76$ TeV and also to make predictions for measurements at $\sqrt{s} = 5.02$ TeV. The observables considered included charged particle multipicity distributions, triple-differential spectra, differential and integrated flow ($v_n \{2 \le n \le  5\}$), averaged transverse momentum,  event-by-event flow coefficients distributions, and measures of event-plane correlation \cite{McDonald:2016vlt}. The results for the calculation of integrated flow coefficients and their comparison with ALICE measurements are show in Figure \ref{int_flow}. The agreement between theory with data shown there uses a hydrodynamic phase with an effective specific shear viscosity coefficient of $\eta/s$ = 0.095, and a temperature-dependent bulk viscosity $\zeta/s (T)$ \cite{McDonald:2016vlt}. The fact that the specific shear viscosity is the same at both LHC energies suggests that its temperature-dependence in the region spanned by the CERN facility is mild. To summarize this section, simulation approaches of relativistic nuclear collisions have attained a certain degree of advancement, as quantified by the agreement between model and a wealth of data \cite{sangwook}. A corresponding level of evolution must now be imposed on electromagnetic variables.

\section{Photons}

\begin{figure*}[h!]
\centering
%\begin{tabular}{ccc}
\includegraphics[width=0.23\linewidth]{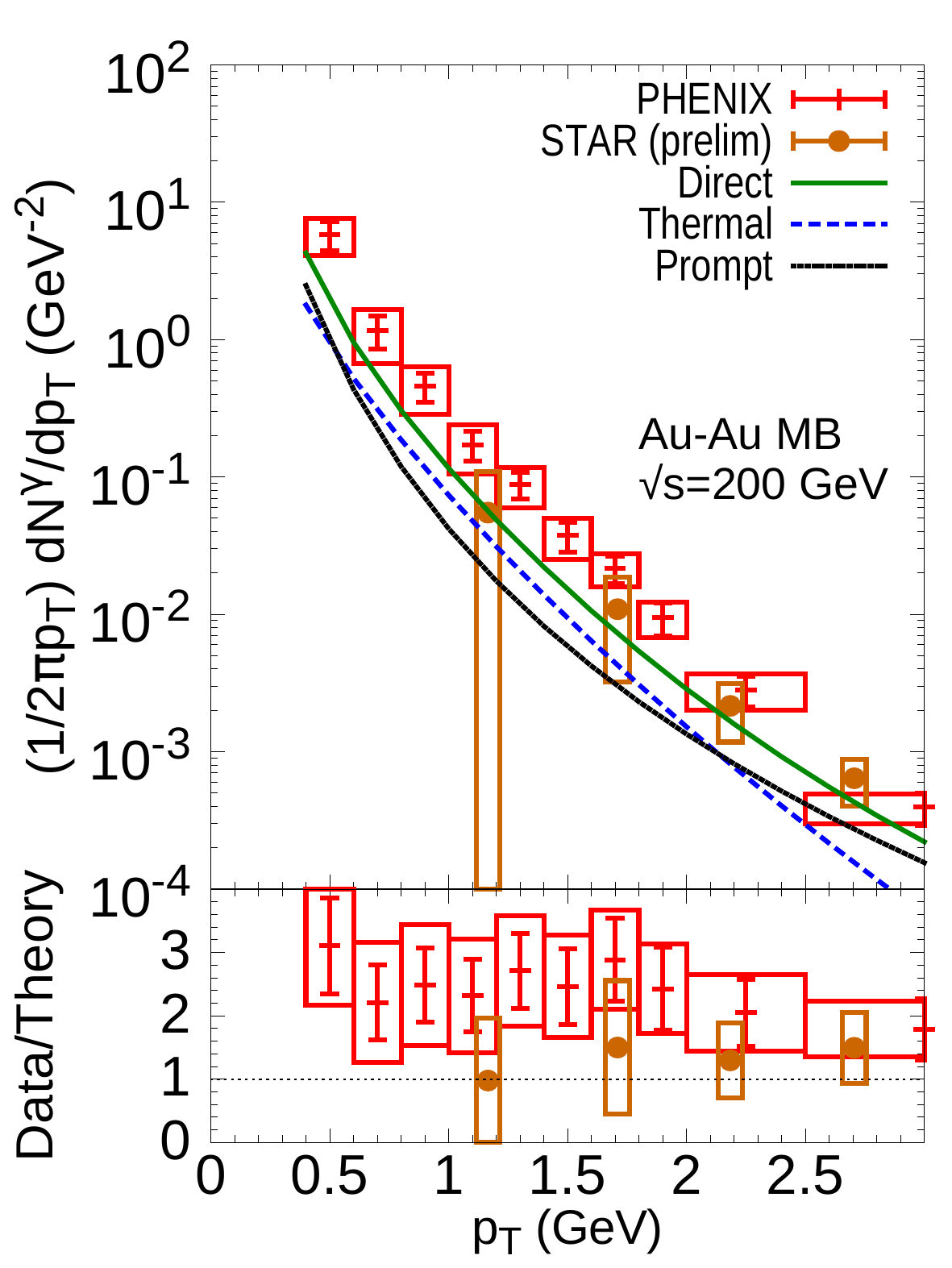}
\includegraphics[width=0.23\linewidth]{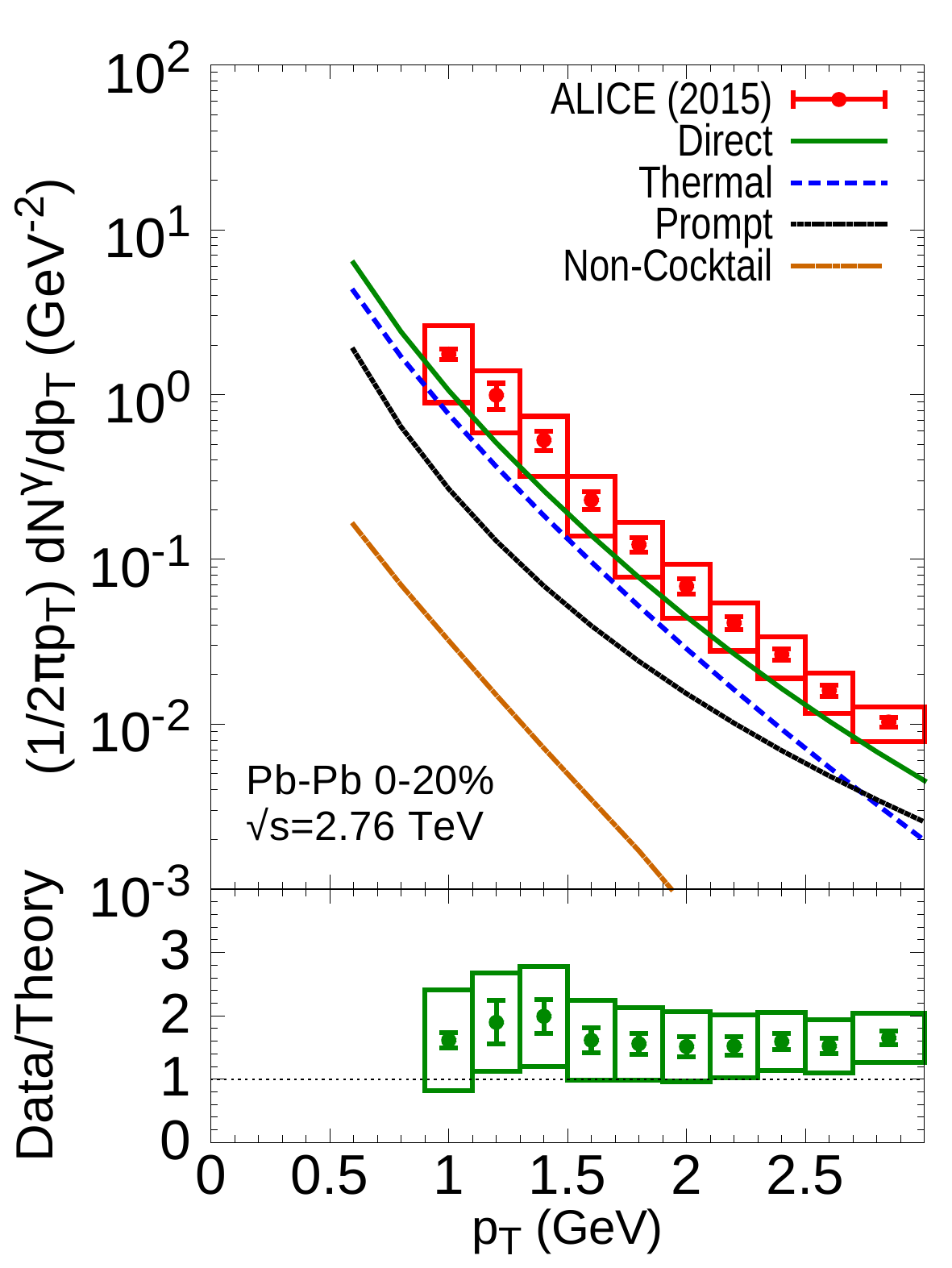}\\\hspace*{0.07cm}
\includegraphics[width=0.23\linewidth]{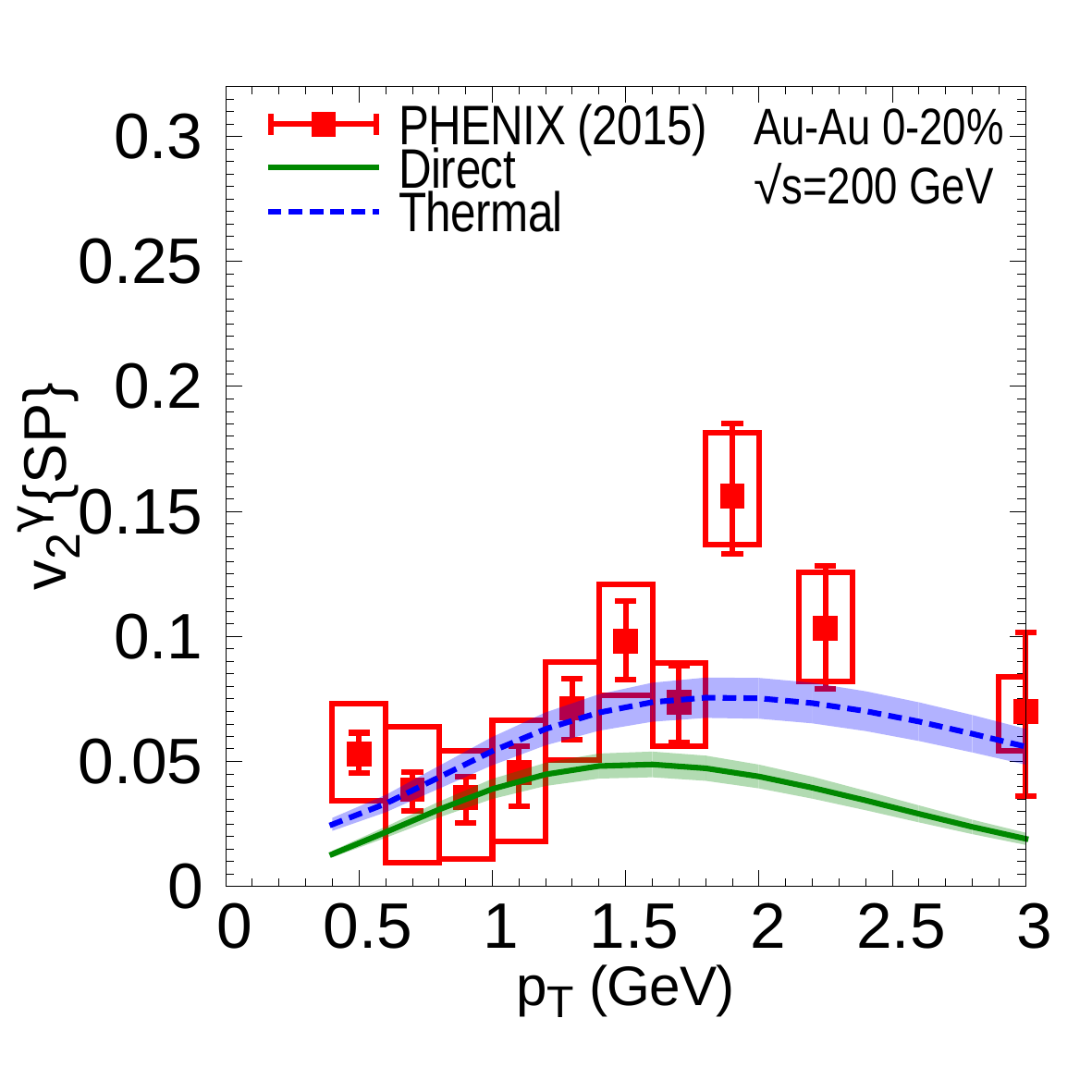}\hspace*{0.04cm}
\includegraphics[width=0.23\linewidth]{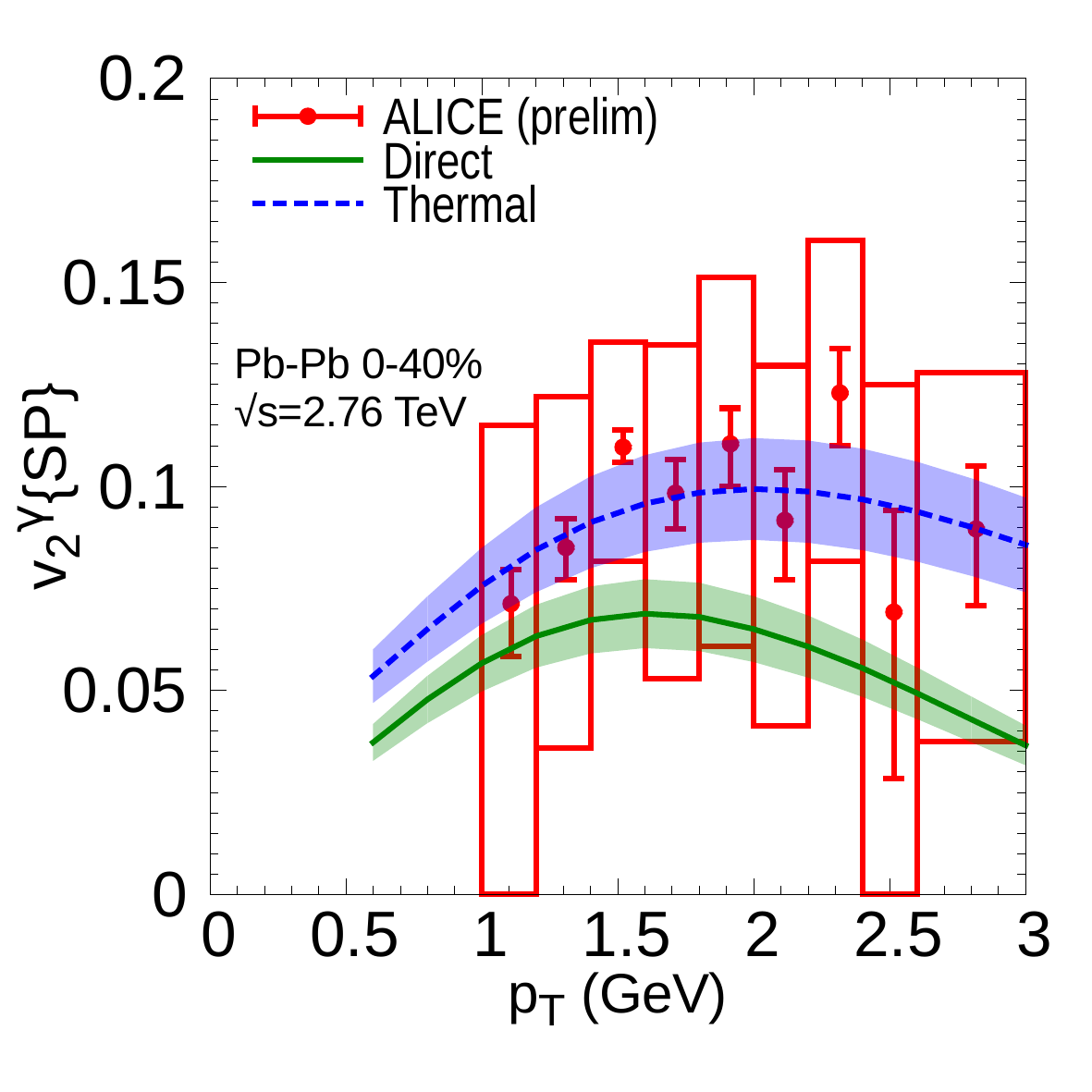}
%  \includegraphics[width=0.40\linewidth]{STAR_photons.pdf} 
%  \includegraphics[width=0.30\linewidth]{Fig1b.pdf} & 
%  \includegraphics[width=0.30\linewidth]{Fig1c.pdf}
%\end{tabular}
\caption{{\it (Top Row) Left Panel}: Inclusive direct photon spectra from the PHENIX and STAR collaborations, together with different contributions of the direct photon signal. {\it Right Panel}: Direct central photon data from ALICE. Linear ratios of data/theory are also shown. {\it (Bottom Row) Left Panel}: Photon elliptic flow, as measured at RHIC (left panel) and at the LHC (right panel). The dotted curves show the thermal signal only, and the solid lines show the net photon elliptic flow. Calculation results are from \cite{Paquet:2015lta}.}
\label{pho-spectra}
\end{figure*}

The very first interactions of the hadronic  entities in collisions will produce ``prompt photons'', calculable using techniques of perturbative QCD \cite{pQCD}. From intermediate to high $p_T$ (to be made more precise later), prompt photons will constitute the dominant component of direct photons.  They are calculated by multiplying the number of photons produced in nucleon-nucleon collisions by the number of binary collisions, accounting for isospin, and taking into consideration  the modification in cold nuclear matter of elementary parton distribution functions. 
Formally, the production cross-section depends on the renormalization, factorization, and fragmentation scales, $Q_{\{x = {\rm R, F, Fr}\}}$. 
%\begin{eqnarray}
%E \frac{d^3 \sigma_{pp}}{d^3 p} = \sum_{a, b, c, d} f_{a/p} (x_a, Q_{\rm fact}) \otimes f_{b/p}( x_b, Q_{\rm fact}) \otimes d \hat\sigma (Q_{\rm ren}) \otimes D_{\gamma/c} (z_c, Q_{\rm frag})
%\end{eqnarray}
%where the $\otimes$ represent integration over the kinematical variables $x, z$ with the appropriate phase space and measure. %The parton distribution functions (p.d.f.), $f$, receive the corrections discussed above, and they depend on a factorization scale $Q_{\rm frag}$. 
There is no formal theoretical guidance on where exactly to set the values of those different scales, but this uncertainty can also be used to one's advantage \cite{JFThesis} and these choices  should ideally be validated against nucleon-nucleon data. These photons constitute an irreducible background for the ones discussed next. 
%The parton-parton differential cross section, $d \sigma$, is to be computed as a given order in the strong coupling that is consistent with that of the p.d.f. Finally the fragmentation function $D$ is a non-perturbative entity, and is extracted from  measurement results. 

The ``thermal photons'' are those photons which result from interactions between thermalized, or approximately thermalized, constituents of the strongly interacting medium. %This  enables a connection with the theoretical treatment of hadrons: the fluid dynamical medium is known to be viscous, and this non-equilibrium feature will affect the characteristics of all measured observables. 
Then, if only for theoretical consistency, the treatment of electromagnetic emission must take into account the non-equilibrium characteristics of the medium. Some work has been done in this direction, but not all photon-producing processes have yet been corrected for viscous effects: this is discussed later in this section.  For now, an appropriate snapshot of the state of the theory appears in Table II of Ref. \cite{Paquet:2015lta}. %Uniting the hydrodynamical modelling discussed previously with state of the art electromagnetic emissivities still leaves tension between experimental data and theory, especially in what concerns photon differential flow ($v_2$ and $v_3$). This has been dubbed the ``photon flow puzzle''. 
A summary of photon spectra and elliptic flow calculated with viscous (shear and bulk) hydrodynamics appears in Fig. \ref{pho-spectra}, together with the data garnered by the PHENIX, STAR, and ALICE collaborations at RHIC and at the LHC. Clearly, there is a tension between theory and photon spectrum data, but especially in what concerns the photon elliptic flow (and more so at RHIC). This has been dubbed the ``photon flow puzzle''\footnote{Updated photon spectra and flow coefficients obtained with model parameters and components  updated from those in Ref. \cite{Paquet:2015lta} are in preparation and will appear elsewhere, together with a fresh perspective on the ``photon flow puzzle''.}. It is fair to write that this tension remains, in spite of many attempts to ameliorate photon emission rates\footnote{Not an exhaustive list.} \cite{Basar:2012bp}.  %,Chiu:2012ij,Muller:2013ila,Gale:2014dfa,Linnyk:2015rco,Greif:2016jeb,Ayala:2017vex,Berges:2017eom}. 
However, another related puzzle has appeared: the STAR Collaboration has measured virtual photons and deduced from them real photon spectra; their inclusive data are also shown in Fig. \ref{pho-spectra}. The STAR spectra are much closer to theoretical results in all centrality classes \cite{STAR:2016use}, but STAR and PHENIX photons are currently incompatible, even considering  uncertainty bounds. This discrepancy begs a resolution, and this is now in fact a prerequisite  for the field to move forward. The STAR Collaboration does not yet have results for photon elliptic flow. 

Going back to more formal aspects, more progress on this topic will require extending the theory of photon emission to regions out of equilibrium. Up to now, the inclusion of viscosity in the electromagnetic emissivity mostly relied on the kinetic theory formulation of the emission rates \cite{Dion:2011pp,Shen:2014nfa,JFThesis,Paquet:2015lta}. However, the Landau-Pomeranchuck-Migdal effect (LPM) is important  for photon emission at leading order in the strong coupling, and it involves an arbitrary number of gluon exchanges; its calculation in equilibrium is performed using techniques of finite-temperature field theory (FTFT) \cite{AMY} that rely on the Kubo-Martin-Schwinger (KMS) relation which is a statement of detailed balance \cite{KG}. A general field-theoretical formalism to evaluate the production of electromagnetic radiation to complete leading order in the strong coupling in a out-of-equilibrium medium had been lacking, but has been derived recently \cite{SiggiMSc}. Its phenomenological consequences are left for future work, but we present here some results that illustrate its clear potential. 

The rate for photon emission from an equilibrium medium \cite{KG} now needs to be generalized, to allow for the consideration of a medium out of equilibrium using real-time finite-temperature field theory \cite{Baier:1997xc}. Technically, 
\begin{eqnarray}
\omega \frac{d^3 R}{d^3 k} = - \frac{1}{(2 \pi)^3} \frac{1}{e^{\beta \omega} - 1} {\rm Im}\, {\Pi^{\rm R}}_\mu^\mu ( \omega, {\bf k}) \Rightarrow \frac{i}{2 (2 \pi)^3} {\Pi_{1 2}}^\mu_\mu (\omega, {\bf k})\ ,
\end{eqnarray}
where $(\omega, {\bf k})$ is the photon energy and three-momentum, respectively. The finite-temperature, retarded photon self-energy is ${\Pi^{\rm R}}^{\mu \nu}$, and an element of the photon two-point  function appears as $\Pi_{1 2}^{\mu \nu}$ in the \{1,~2\} real-time basis. %These two rate equations use the conventions germane to their specific formalism. 
However, calculations are more economically performed in the $r/a$ basis which also allows for easier power counting: $\phi_r = \frac{1}{2} \left(\phi_1 + \phi_2\right), \phi_a = \phi_1 - \phi_2$. The photo production processes in the medium at leading order in the strong coupling comprises bremsstrahlung, Compton, quark-antiquark annihilation, and LPM  contributions. 
\begin{figure*}[h!]
\centering
%\begin{tabular}{ccc}
  \includegraphics[width=0.350\linewidth]{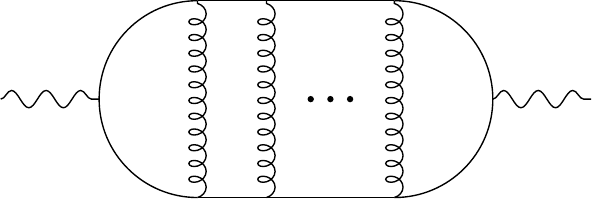} \hspace*{1.5cm}
   \includegraphics[width=0.350\linewidth]{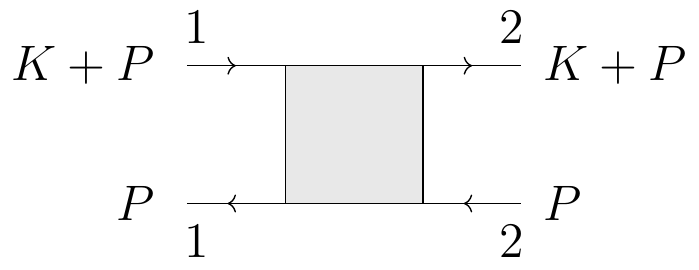} 
%  \includegraphics[width=0.30\linewidth]{Fig1b.pdf} & 
%  \includegraphics[width=0.30\linewidth]{Fig1c.pdf}
%\end{tabular}
\caption{{\it Left Panel:} The self-energy of the photon containing multiple insertions of a soft gluon exchange, which is responsible for the LPM effect. {\it Right panel:} The four-point function $S_{1122}$ in momentum space. Here, $K$ is the photon four-momentum and $P$ is the loop four-momentum. The figures are from Ref. \cite{Hauksson:2017udm}.}
\label{LPM}
\end{figure*}
The photon self-energy diagram that contains the LPM effect\footnote{It also contains bremsstrahlung and pair annihilation contributions.} is shown in Fig. \ref{LPM}, and describes multiple exchanges of soft gluons, whose propagators thus need resummation. The building blocks of this resummation is the fermionic four-point function: $S_{1122} (x_1, x_2; y_1, y_2) = \langle T_c \left\{ \bar\psi_1 (x_1) \psi_1 (x_2) \bar\psi_2 (y_1) \psi_2 (y_2)\right\}\rangle$. In terms of the $r/a$ basis
\begin{eqnarray}
S_{1122} = &S_{rrrr} + \frac{1}{2} \left(S_{arrr} + S_{rarr} - S_{rrar} - S_{rrra}\right)  + \frac{1}{4} \left(S_{aarr}- S_{arar} - S_{arra} -S_{raar} - S_{rara} + S_{rraa}\right)\nonumber \\
& + \frac{1}{8} \left(S_{raaa} + S_{araa} - S_{aara} - S_{aaar} + \frac{1}{2} S_{aaaa}\right)
\end{eqnarray}
This expansion looks formidable, as each term on the right-hand-side hides an infinite sum of diagrams, each with a different number of soft gluon rungs. However, at leading order and in thermal equilibrium it turns out that $S_{1122} = 2 f_F\left(p^0 + k^0\right)\left( 1 - f_F (p^0)\right) {\rm Re}\, S_{aarr}$, where $f_F$ is a Fermi-Dirac distribution function. Away from equilibrium, however, a careful analysis of the topology of the Feynman diagrams contributing to the four-point functions shows that \cite{SiggiMSc,Hauksson:2017udm} (at leading order)
\begin{eqnarray}
&S_{rarr} = \left(\frac{1}{2} - F(P)\right) S_{aarr}, \ S_{rrar} = - \left( \frac{1}{2} - F(P) \right) S_{rraa},\ S_{arrr} = - \left( \frac{1}{2} - F(P + K)\right) S_{aarr},\nonumber\\ 
&S_{rrra} = \left( \frac{1}{2} - F(P+K)\right) S_{rraa}, \ S_{rrrr} = - \left(\frac{1}{2} - F(P)\right) \left( \frac{1}{2} - F(P+K)\right) \left[S_{rraa} + S_{aarr}\right]
\end{eqnarray}
A little more analysis finally shows that $S_{1122} = 2 F (P+K) \left(1 - F(P)\right) \, {\rm Re}\, S_{aarr}$, a result obtained without resorting to the KMS condition. The summation of diagrams implied in $S_{aarr}$ can also be written as an integral equation, as in the case of equilibrium. Note that  $F(P) = f_q (\mathbf{ p}) \theta (p^0) + \left(1 - f_{\bar q}( - {\mathbf p}) \right) \theta(- p^0)$. The function $F(P)$ is thus to be interpreted as an occupation density: in the Boltzmann equation, incoming particles have $p^0 > 0$ and outgoing particles $p^0 < 0$. For outgoing quarks, $F(P)$ is thus simply the bare momentum distribution, with the Pauli blocking correction. In equilibrium, it reduces to a Fermi-Dirac distribution in equilibrium, since $1 - f_F (-x) = f_F(x)$. 

Putting the elements together, one gets the rates for real photon emission from a non-equilibrium medium to be \cite{SiggiMSc,Hauksson:2017udm}
\begin{eqnarray}
\omega \frac{d^3R}{d^3 k} = \frac{3 Q^2 \alpha_{\rm EM}}{4 \pi^2} \int \frac{d^3 p}{(2 \pi)^3} F(P + K) \left[1 - F(P)\right] \frac{p_z^2 + (p_z + k^2)^2}{2 p_z^2 (p_z + k)^2}\, {\mathbf p}_\perp \cdot {\rm Re}\, {\mathbf f}({\mathbf p}; {\mathbf k})\ ,
\label{rate}
\end{eqnarray}
where $Q$ is obtained by summing over charges of light quark flavours (in units of the electron charge), and  ${\mathbf f}$ satisfies an integral equation of the Boltzmann-type with a collision kernel ${\mathcal C}$:
\begin{eqnarray}
{\mathbf p}_\perp = i  \delta E\, {\mathbf f}({\mathbf p}_\perp) + \int \frac{d^2 q_\perp}{(2 \pi)^2} {\mathcal C} ({\mathbf q}_\perp ) \left[ {\mathbf f}({\mathbf p}_\perp) - {\mathbf f}( {\mathbf p}_\perp + {\mathbf q}_\perp\right )]
\end{eqnarray}
For the special case of an isotropic plasma, $f ( {\mathbf p} ) = f( p )$, one recovers a collision kernel that can be written as \cite{Arnold:2002zm}
\begin{eqnarray}
{\mathcal C} ({\mathbf q}_\perp) = g^2 C_F \Omega \left[ \frac{1}{{\mathbf q}_\perp^2} - \frac{1}{{\mathbf q}_\perp^2 + m_{\rm D}^2}\right]\ ,
\end{eqnarray}
with $C_F$ the Casimir operator, $m_{\rm D}^2 $  the non-equilibrium Debye mass, and $\Omega$ characterizing the occupation density of soft gluons \cite{SiggiMSc,Hauksson:2017udm}.
%\begin{eqnarray}
%\Omega = \frac{\int_0^\infty d p p^2 \left[2 N_f f_q \left( 1 - f_q\right) + 2 N_c f_g \left( 1 + f_g \right)\right]}{- \int_0^\infty d p p^2 \frac{d}{d p} \left[2 N_f f_q + 2 N_c f_g \right]}\ ,
%\end{eqnarray}
%and where $m_{\rm D}^2 $ is the non-equilibrium Debye mass. % is
%\begin{eqnarray}
%m_{\rm D}^2 = \frac{g^2}{\pi^2} \int_0^\infty d p p^2 \left[ 2 N_f f_q (p) + 2 N_c f_g (p)\right]
%\end{eqnarray}
In an anisotropic plasma, the collision kernel ${\mathcal C} ({\mathbf q}_\perp )$ and its integral, the quark decay width, are formally divergent, owing to gauge field instabilities \cite{Mrowczynski:2016etf}. The analysis performed here  assumes implicitly that the plasma anisotropy is small enough for the divergences not to appear at leading order in the strong coupling. An exploration of the  associated  dynamical restrictions will appear elsewhere, together with solutions of Eq. (\ref{rate}), and its phenomenological consequences on the soft photon puzzle and on other related issues. 

\section{Conclusions}
Significant progress  has been accomplished in the theory of photon production in relativistic nuclear collisions and in the modelling of heavy-ion collision dynamics. Several unknowns remain, but we have reported on a systematic study of leading-order out-of-equilibrium photon production which should put the computation of electromagnetic emissivity in a variety of environments  on a firm footing. 

\section*{Acknowledgements}
\noindent I am happy to thank Sigtryggur Hauksson for a critical reading of this manuscript. %, and I acknowledge students, postdocs, and collaborators with whom much of the work reported on here was done. 
I am grateful for the support of the Canada Council for the Arts -- through its Killam Research Fellowships Program -- and for that of the Local Organizing Committee.  This work was funded in part by the Natural Sciences and Engineering Research Council of Canada. 

\


\begin{thebibliography}{99}
%\bibitem{Bjorken:1982tu} 
%  J.~D.~Bjorken,
  %``Energy Loss of Energetic Partons in Quark - Gluon Plasma: Possible Extinction of High p(t) Jets in Hadron - Hadron Collisions,''
%  FERMILAB-PUB-82-059-THY, FERMILAB-PUB-82-059-T.
  %%CITATION = FERMILAB-PUB-82-059-THY, FERMILAB-PUB-82-059-T;%%
  %246 citations counted in INSPIRE as of 29 Jan 2018
%\bibitem{Connors:2017ptx} 
%See, for example,   M.~Connors, C.~Nattrass, R.~Reed and S.~Salur,
  %``Review of Jet Measurements in Heavy Ion Collisions,''
%  arXiv:1705.01974 [nucl-ex], and references therein.
  %%CITATION = ARXIV:1705.01974;%%
  %8 citations counted in INSPIRE as of 29 Jan 2018
\bibitem{Gale:2009gc} 
See, for example,   C.~Gale,
  %``Photon Production in Hot and Dense Strongly Interacting Matter,''
  Landolt-Bornstein {\bf 23}, 445 (2010), and references therein.

\bibitem{Gale:2013da} 
See, for example, C.~Gale, S.~Jeon and B.~Schenke,
  %``Hydrodynamic Modeling of Heavy-Ion Collisions,''
  Int.\ J.\ Mod.\ Phys.\ A {\bf 28}, 1340011 (2013), and references therein.
  %%CITATION = doi:10.1142/S0217751X13400113;%%
  %310 citations counted in INSPIRE as of 15 Jan 2018
\bibitem{Poskanzer:1998yz} 
  A.~M.~Poskanzer and S.~A.~Voloshin,
  %``Methods for analyzing anisotropic flow in relativistic nuclear collisions,''
  Phys.\ Rev.\ C {\bf 58}, 1671 (1998).
\bibitem{Luzum:2013yya} 
  M.~Luzum and H.~Petersen,
  %``Initial State Fluctuations and Final State Correlations in Relativistic Heavy-Ion Collisions,''
  J.\ Phys.\ G {\bf 41}, 063102 (2014).
  \bibitem{Jeon:2015dfa} 
  S.~Jeon and U.~Heinz,
  %``Introduction to Hydrodynamics,''
  Int.\ J.\ Mod.\ Phys.\ E {\bf 24}, no. 10, 1530010 (2015).
\bibitem{McDonald:2016vlt} 
  S.~McDonald, C.~Shen, F.~Fillion-Gourdeau, S.~Jeon and C.~Gale,
  %``Hydrodynamic predictions for Pb+Pb collisions at 5.02 TeV,''
  Phys.\ Rev.\ C {\bf 95}, no. 6, 064913 (2017).
\bibitem{IPGlasma} 
  B.~Schenke, P.~Tribedy and R.~Venugopalan,
  %``Event-by-event gluon multiplicity, energy density, and eccentricities in ultrarelativistic heavy-ion collisions,''
  Phys.\ Rev.\ C {\bf 86}, 034908 (2012); {\it idem}, 
  %``Fluctuating Glasma initial conditions and flow in heavy ion collisions,''
  Phys.\ Rev.\ Lett.\  {\bf 108}, 252301 (2012).
\bibitem{ALICE_flow}
J.~Adam {\it et al.} [ALICE Collaboration],
  %``Anisotropic flow of charged particles in Pb-Pb collisions at $\sqrt{s_{\rm NN}}=5.02$ TeV,''
  Phys.\ Rev.\ Lett.\  {\bf 116}, no. 13, 132302 (2016); 
  K.~Aamodt {\it et al.} [ALICE Collaboration],
  %``Higher harmonic anisotropic flow measurements of charged particles in Pb-Pb collisions at $\sqrt{s_{NN}}$=2.76 TeV,''
  Phys.\ Rev.\ Lett.\  {\bf 107}, 032301 (2011).
%  \bibitem{music_web}
%  http://www.physics.mcgill.ca/music/
  \bibitem{sangwook}
  See also S. Ryu {\it et al.}, these proceedings.
 \bibitem{pQCD}
 See, for example, J. F. Owens, Rev. Mod. Phys. {\bf 59}, 465 (1987), for an early review. More recent analyses can be found in  M.~Klasen, C.~Klein-B\"osing, F.~K\"onig and J.~P.~Wessels,
  JHEP {\bf 1310}, 119 (2013);   %M.~Klasen, F.~K\"onig, C.~Klein-B\"osing and J.~P.~Wessels,
  Nucl.\ Part.\ Phys.\ Proc.\  {\bf 273-275}, 1509 (2016).
 \bibitem{JFThesis} Jean-Fran\c{c}ois Paquet, Ph.D. thesis, McGill University, 2015.
 \bibitem{Paquet:2015lta} 
 J.~F.~Paquet, C.~Shen, G.~S.~Denicol, M.~Luzum, B.~Schenke, S.~Jeon and C.~Gale,
  Phys.\ Rev.\ C {\bf 93}, no. 4, 044906 (2016). 
\bibitem{Basar:2012bp} 
  G.~Basar, D.~Kharzeev, D.~Kharzeev and V.~Skokov,
  %``Conformal anomaly as a source of soft photons in heavy ion collisions,''
  Phys.\ Rev.\ Lett.\  {\bf 109}, 202303 (2012); 
  %\bibitem{Chiu:2012ij} 
  M.~Chiu, T.~K.~Hemmick, V.~Khachatryan, A.~Leonidov, J.~Liao and L.~McLerran,
  %``Production of Photons and Dileptons in the Glasma,''
  Nucl.\ Phys.\ A {\bf 900}, 16 (2013);
%  \bibitem{Muller:2013ila} 
  B.~Muller, S.~Y.~Wu and D.~L.~Yang,
  %``Elliptic flow from thermal photons with magnetic field in holography,''
  Phys.\ Rev.\ D {\bf 89}, no. 2, 026013 (2014);
%  \bibitem{Gale:2014dfa} 
  C.~Gale {\it et al.},
  %``Production and Elliptic Flow of Dileptons and Photons in a Matrix Model of the Quark-Gluon Plasma,''
  Phys.\ Rev.\ Lett.\  {\bf 114}, 072301 (2015); 
%\bibitem{Linnyk:2015rco} 
  O.~Linnyk, E.~L.~Bratkovskaya and W.~Cassing,
  %``Effective QCD and transport description of dilepton and photon production in heavy-ion collisions and elementary processes,''
  Prog.\ Part.\ Nucl.\ Phys.\  {\bf 87}, 50 (2016); 
%\bibitem{Greif:2016jeb} 
  M.~Greif, F.~Senzel, H.~Kremer, K.~Zhou, C.~Greiner and Z.~Xu,
  %``Nonequilibrium photon production in partonic transport simulations,''
  Phys.\ Rev.\ C {\bf 95}, no. 5, 054903 (2017); 
%\bibitem{Ayala:2017vex} 
  A.~Ayala, these proceedings;  %, J.~D.~Castano-Yepes, C.~A.~Dominguez, L.~A.~Hernandez, S.~Hernandez-Ortiz and M.~E.~Tejeda-Yeomans,
  %``Prompt photon yield and elliptic flow from gluon fusion induced by magnetic fields in relativistic heavy-ion collisions,''
%  Phys.\ Rev.\ D {\bf 96}, no. 1, 014023 (2017)
%  Erratum: [Phys.\ Rev.\ D {\bf 96}, no. 11, 119901 (2017)].
%\bibitem{Berges:2017eom} 
  J.~Berges, K.~Reygers, N.~Tanji and R.~Venugopalan,
  %``Parametric estimate of the relative photon yields from the glasma and the quark-gluon plasma in heavy-ion collisions,''
  Phys.\ Rev.\ C {\bf 95}, no. 5, 054904 (2017).
 
 
 
  \bibitem{STAR:2016use} 
  L.~Adamczyk {\it et al.} [STAR Collaboration],
  %``Direct virtual photon production in Au+Au collisions at $\sqrt{s_{NN}}$ = 200 GeV,''
  Phys.\ Lett.\ B {\bf 770}, 451 (2017). 
 
 
\bibitem{Dion:2011pp} 
Maxime Dion, M. Sc. thesis, McGill University, 2011; 
  M.~Dion, J.~F.~Paquet, B.~Schenke, C.~Young, S.~Jeon and C.~Gale,
  %``Viscous photons in relativistic heavy ion collisions,''
  Phys.\ Rev.\ C {\bf 84}, 064901 (2011).

\bibitem{Shen:2014nfa} 
  C.~Shen, J.~F.~Paquet, U.~Heinz and C.~Gale,
  %``Photon Emission from a Momentum Anisotropic Quark-Gluon Plasma,''
  Phys.\ Rev.\ C {\bf 91}, no. 1, 014908 (2015).  
  \bibitem{AMY}
  Peter Brockway Arnold, Guy D. Moore, and Laurence G. Yaffe, JHEP {\bf 11}, 057 (2011); {\bf 12}, 009 (2001).
\bibitem{KG}
Joseph I. Kapusta and Charles Gale, {\it Finite-Temperature Field Theory: Principles and Applications}, Cambridge University Press, 2006.
\bibitem{SiggiMSc}Sigtryggur Hauksson, M. Sc. thesis, McGill University, 2017.
\bibitem{Baier:1997xc} 
  R.~Baier, M.~Dirks, K.~Redlich and D.~Schiff,
  %``Thermal photon production rate from nonequilibrium quantum field theory,''
  Phys.\ Rev.\ D {\bf 56}, 2548 (1997).
  \bibitem{Hauksson:2017udm} 
  S.~Hauksson, S.~Jeon and C.~Gale,
  %``Photon emission from quark-gluon plasma out of equilibrium,''
  Phys.\ Rev.\ C {\bf 97}, no. 1, 014901 (2018).
\bibitem{Arnold:2002zm}
 P.~B.~Arnold, G.~D.~Moore and L.~G.~Yaffe,
  %``Effective kinetic theory for high temperature gauge theories,''
  JHEP {\bf 0301}, 030 (2003)
\bibitem{Mrowczynski:2016etf} 
  S.~Mrowczynski, B.~Schenke and M.~Strickland,
  %``Color instabilities in the quarkÐgluon plasma,''
  Phys.\ Rept.\  {\bf 682}, 1 (2017).
  \end{thebibliography}
\end{document}